Review

Qihang Ai, Hanxiao Feng, Xinyu Yang, Mengxi Tan*, Xingyuan Xu, Roberto Morandotti, Donglin Su*, and David J. Moss*

# Photonic real-time signal processing

**Abstract:** The simultaneous progress of integrated optical frequency comb (OFC) and radio frequency (RF) photonic signal processing technique have promoted the rapid development of real-time signal processing. Integrated optical frequency comb offer multiple wavelengths as a powerful source for RF photonic signal transversal filter. Here, we review development of real-time signal processing system consisting of integrated OFC and RF photonic signal transversal filter in chronological order, and focus on the applications of this system such as differentiator, integrator, Hilbert transformer, and image processor. We also discuss and present our outlook on more parallel functions and further integration of real-time signal processing system.

**Keywords:** RF photonics; optical frequency comb; real-time signal processing system

**Qihang Ai,** School of Electronic and Information Engineering, Beihang University, Beijing 100191, China. e-mail: aiqihang@buaa.edu.cn; ORCID
**Hanxiao Feng,** School of Electronic and Information Engineering, Beihang University, Beijing 100191, China. e-mail: fenghanxiao@buaa.edu.cn; ORCID
**Xinyu Yang,** School of Electronic and Information Engineering, Beihang University, Beijing 100191, China. e-mail: xinyu_yang1022@hotmail.com; ORCID
**\*Corresponding author: Mengxi Tan,** School of Electronic and Information Engineering, Beihang University, Beijing 100191, China. e-mail: simtan@buaa.edu.cn; ORCID
**Xingyuan Xu,** State Key Laboratory of Information Photonics and Optical Communications, Beijing University of Posts and Telecommunications, Beijing 100876, China. e-mail: xingyuanxu@bupt.edu.cn; ORCID
**Roberto Morandotti,** INRS-Énergie, Matériaux et Télécommunications, 1650 Boulevard Lionel-Boulet, Varennes, Québec J3X 1S2, Canada. e-mail: roberto.morandotti@inrs.ca; ORCID
**\*Corresponding author: Donglin Su,** School of Electronic and Information Engineering, Beihang University, Beijing 100191, China. e-mail: sdl@buaa.edu.cn; ORCID
**\*Corresponding author: David J. Moss,** Optical Sciences Centre, Swinburne University of Technology, Hawthorn, VIC 3122, Australia., ARC Centre of Excellence in Optical Microcombs for Breakthrough Science, Australia., e-mail: dmoss@swin.edu.au; ORCID

## 1 Introduction

With the rapid development of the technology, information density has increased. The speed of information transmission and processing has been restricted by the electronic bottleneck defined by moore's Law, which makes it difficult to meet the needs of information age. Compared with conventional real-time signal processing technology, RF photonics can modulate real-time signal into the optical domain which can be transmitted and processed by optical technology and further realize high speed and wide bandwidth [1–5].

RF photonics is an interdisciplinary technology that combines RF and photonic technology, which can utilize the advantage of broad bandwidth, ultrahigh-speed and low loss offered by photonics to generate, transmit and process real-time signal [6, 7]. RF photonics can be used in radar systems, wireless communication, radio frequency identification, biomedical imaging, astronomy and quantum information processing [8–10]. Conventional real-time signal processing technologies typically operate in frequency range from tens of kilohertz to tens of gigahertz for processing low frequency real-time signals and the speed of conventional real-time signal processing technology is limited by electronic devices [11–13]. Compared with conventional real-time signal processing technology, the real-time signal processing based RF photonics can operate in frequency range from hundreds of megahertz to hundreds of terahertz for processing higher frequency real-time signal in higher speed. Real-time signal processing based RF photonics utilize light to transmit real-time signal and then have greater resilience to electromagnetic interference resulting in higher signal quality [14, 15].

Optical frequency comb (OFC) is a multiple-wavelength light source that offers numerous coherent wavelength channels. The adjacent channels are equally spaced in the OFC spectrum. Conventional OFC generation approaches feature large size, high cost, and limited wavelengths [16–19]. By contrast, micro-ring resonators with ultra-high Q factor have a point that offers a wide bandwidth in a little space, and can serve as an ideal platform for OFC generation [20–22]. In 2015, OFC based on a micro-ring resonator was first applied to RF photonics for real-time ultra-fast signal processing, establishing the connection between optical frequency and RF frequency [23].

## 2 RF Photonic Signal Processing

The RF photonic signal processing is a rapidly developing technology which has been widely used in many fields such as image processing, telecommunication, photon radar and photon neural network [8–10]. Compared to the conventional RF link filters, the RF signal processing system has higher bandwidth up to THz, lower loss about 0.15 dB/km and more suitable for long-distance transmission [13]. It transforms the electrical signals into optical signals to use optical methods to process signals then transform the optical signals back to the electrical signals as an output. A conventional RF photonic signal processing system contains light source, electro-optical conversion, optical signal processing unit and photoelectric conversion. Fig. 2 is the model of the typical RF photonic signal processing system. For example a RF photonic signal processing system first generate a Kerr combs then shaping the combs which is the impluse response of the system in order to obtain the required system functions then modulating the RF signal through the Mach-Zehnder modulator (MZM). After modulating the RF signal the system uses the time-delay fiber to Implement convolution and finally transform the optical signals back to the electrical signal as an output. The MZM has two optical paths, a RF signal input and a bias voltage input which can transform the phase information into intensity information with the advantages of fast speed, good thermal stability, high extinction ratio, and low chirp [66]. The optical signal is equally distributed across two optical paths according to power and the RF input signal is loaded into two optical paths. Under the effect of the electric field, the propagation speed of light in an optical waveguide will change, resulting in phase difference so that the RF signal could be modulated into the optical signal. The transfer function of the MZM is a cosine function, so it is necessary to adjust the bias voltage to ensure that the modulator operates in the linear region.

The demonstration of RF photonic signal Processing system could be date back to 1977 by Chang [68]. He used 15 multi-mode fibers to create 15 different intervals with a delay of 5.2 ns demonstrating a filter which has 193 MHz fundamental passband. Later in 1995 Lindsay demonstrated the 20 GHz bandwidth photonic mixers and applied to superheterodyne receivers [34]. Frankel used 8 different fibers demonstrating a 8 taps tunable filter and this system used a low voltage control to change the wavelength of the laser continuously which enable the filter to continuously change the passband [24]. Another research of the mixer is demonstrated by Michael who extended the frequency band range to 1 THz [35]. Yongwoo Park used the similar structure to demonstrate the photonic intensity high order temporal differentiator in 2009 [47]. He used the variable attenuator to change the amplitude of the taps which is more reconfigurable and flexible. Junqiang Zhou also proposed a $LiNbO_3$ phase modulator photonic differentiator [46]. The more complex RF photonic signal processing system used a tunable laser source to connect a set of 1×8 Bragg gratings through a power divider, each grating equipped with an attenuator to achieve different tap weights, and thus demonstrating an 8-tap filter [69]. However, filters based on Bragg gratings have the disadvantages of narrow bandwidth and poor low-frequency cutoff performance, which is because it is difficult to manufacture gratings that strictly meet the requirements [70–72]. The RF photonic signal processing system also applied in the neural network. In 2012 Lager and Paquot used intensity encoding calculation to implement a photonic neural network which had faster computing speed than the traditional neural network [51]. At the same time, the photonic mixer continued enhancing their dynamic range to 127 dB·$Hz^{4/5}$ [38] and the conversion efficiency to 11.3 dB [36] through the Brillouin scattering and the dual-parallel Mach-Zehnder modulator (DPMZM). The photonic filters, the photonic Hilbert transformer and the photonic neural network also developed at the direction of higher bandwidth, higher integration by the use of the DPMZM [61], the brillouin scattering [26, 28], the crystal delay line [27] and the ring resonator [58].

Thach used the RF photonic signal Processing system based on the kerr combs to demonstrate a Hilbert transformer with impulse response in hyperbolic cosine function which has 20 taps and a wide 3 dB bandwidth above 5 octave [63]. After the demonstration of the 20 taps Hilbert transformer, Xingyuan Xu demonstrated a 80 taps microwave photonic bandpass filter [33]. The record number of tap enabled the filters to achieve excellent performance which was the Four times higher than the previous Q value, 0.5 GHz-4.6 GHz 3 dB bandwidth and 48.9 dB out of band inhibition. The experimental data was highly consistent with the simulation results, indicating that this method is effective. Moss further improved the performance of the Hilbert converter. In addition, the Kerr comb also was applied in the integrator [60] the differentiator [49, 50] and the neural network [49, 73]. Along with the application of soliton crystal frequency combs in neural networks, research had also been conducted on the training [53] and reinforcement learning of photonic neural networks [52]. Futhermore, multiplication and accumulation operations had been achieved by using photonic neural networks [54]. There are also some filters based on the electric optical frequency comb [25].

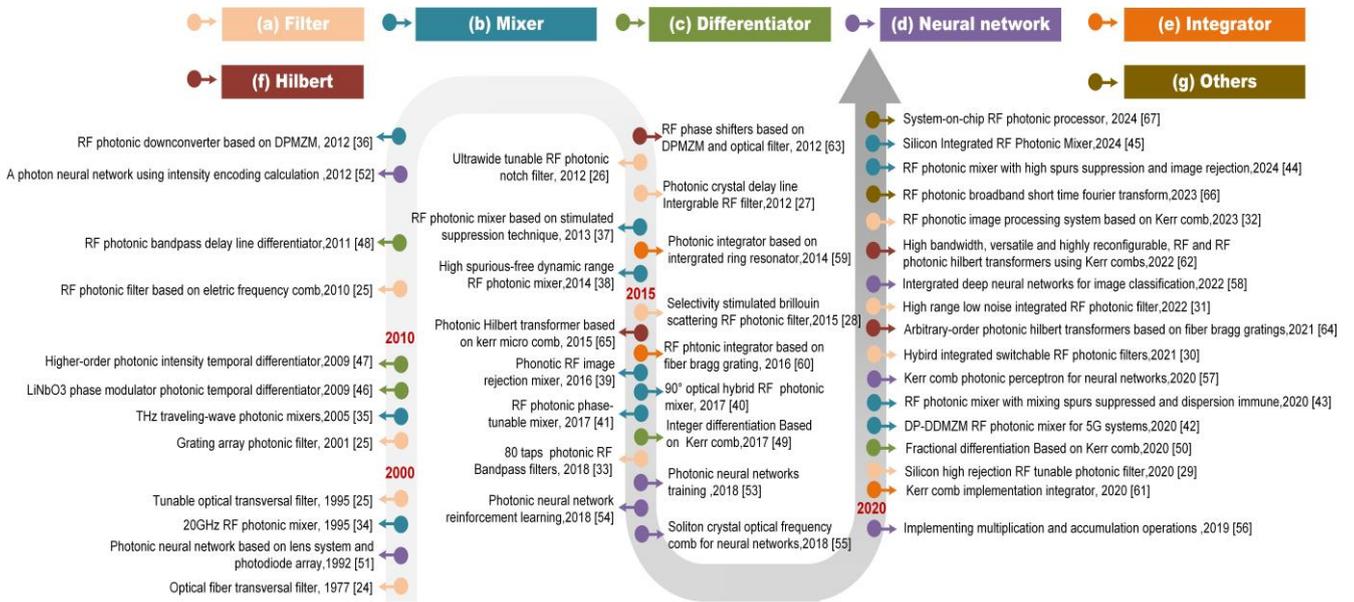

**Fig. 1:** The development of the photonic signal processing including. (a) Filters [24–33]. (b) Mixers [34–39, 39–45]. (c) Differentiators [46–50]. (d) Neural networks [49, 51–57]. (e) Integrators [58–60]. (f) Hilbert transformers [61–63]. (g) Others [64, 65].

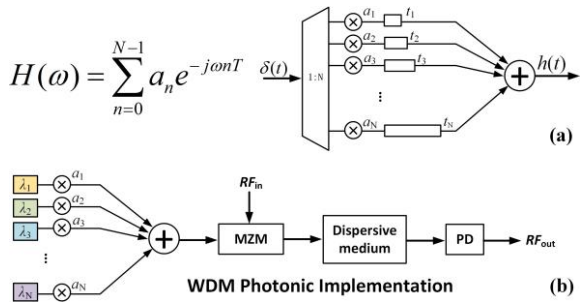

**Fig. 2:** (a) The structure of the RF signal processing system. (b) The actual composition of the system, weight coefficient is provided by the the length of the optical frequency comb, the signal mode fiber provide the different time-delay depending on the different wavelength and then calculate them all by the photodetector [67]. PD: photodetector; MZM: Mach-Zehnder modulator.

After 2020s, the RF photonic signal processing system are developing in the direction of silion intergrated [29, 30, 45, 57, 65] lower noise [32, 43, 44] and higher bandwidth [64]. Due to the advantages of high processing speed, integration, low loss, and large bandwidth, photonic signal processing systems have broad research prospects. For example, Mengxi Tan applied a RF photonic signal processing system based on Kerr comb in the field of image processing. This system achieved high reconfigurability and had the potential for integration [74]. The signal processing speed of the system reached an astonishing 17 TBit/s, and it can simultaneously perform 34 image processing functions on about 400,000 video signals. Currently, the integration methods of the RF photonic signal processing system still need further research.

## 3 Optical frequency comb

Optical frequency combs (OFCs) comprise discrete and equidistant phase-locked frequency components, which can cover a range of hundreds THz [92]. The OFCs' time domain is showcased in Fig. 3 (a), appearing as an ultra-short optical pulse sequence. The frequency domain spectrum in Fig. 3 (b) owns the shape of a comb and is obtained through the Fourier transform of the temporal waveform. The OFCs' appearance provides a bridge between optical and RF, allowing the precise measurement of the laser frequency through the electrical means of the RF [93]. The importance of the OFCs was recognized and the scientists, T.W. Hänsch and J. Hall, who made vital breakthroughs in the OFCs field were awarded the 2005 Nobel Prize in Physics. With an increasingly deeper understanding of the OFCs, they attracted enormous attention for various applications in RF photonics [92, 94].

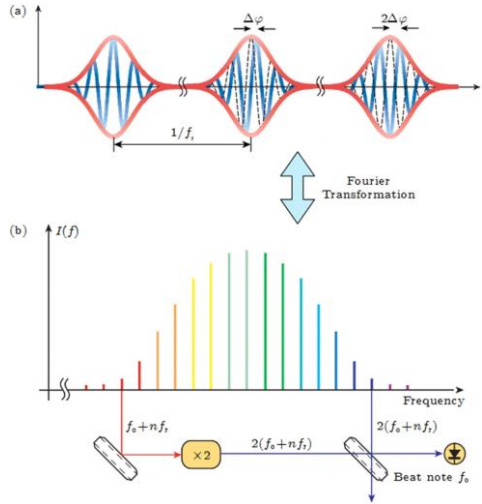

**Fig. 3:** (a) The temporal waveform of the OFC. (b) The optical spectrum of OFC and the two plots have a Fourier relationship [75].

The techniques for generating OFCs include electro-optical modulation, mode-locked laser, or laser array, which result in bulky system size and limited bandwidth [95–100]. To address the limitations, the researcher explored new platforms that produce small-sized and broad bandwidth combs. The development of microcavity with an ultra-high quality factor (Q factor) enables the Kerr microcomb to be a crucial alternative to OFC technology [20, 75, 101].

The discovery of the Kerr microcomb could be traced back to the beginning of the 21st century when Ranka generated the supercontinuum spectrum in microstructured fibers [80]. These materials own anomalous dispersion that stimulates non-linear optical effects such as four-wave mixing and soliton propagation, which are the formation basis for the Kerr combs [80, 82, 102]. In 2003, the silicon-based microcavity with the $10^8$ Q factor was fabricated, in which the optical parameter oscillation was observed in the following year [76–79]. Although only few optical teeth were produced, the results saw the beginning of the Kerr microcombs. Then in 2007, Kippenberg's group pro-duced broadband OFCs in microcavity using the continuous wave (CW) laser, indicating an important node of the microcomb development, after which researchers started to generate Kerr comb in different cavities except for silicon ones, including CMOS-compatible platforms based on Silicon Nitride or high-index doped silica glass [20, 22, 81]. From 2014 to 2019, an increasing number of Kerr comb generation works were conducted, and various soliton combs were observed, namely dissipative Kerr solitons (DKSs), soliton crystal (SCs), and laser cavity solitons (LCSs) [82–84]. The appearance of these low-power consumption and chip-scaled microcombs contributed to the wide application in various fields including RF photon signal processing [85, 86], optical communication [73, 87, 88], precision measurement [89, 90], neural networks [103, 104], and spectroscopy [91], and has a promising future in both the academic and industrial world.

## 4 Real-time signal processing based on OFC

There are many approaches to realize RF photonic signal processing. One is that the optical filter response is mapped onto the RF domain [115–120] and Brillouin scattering by on-chip is the one of the most important approaches [117–119]. Another approach is reconfigurable transfer function for adaptive signal processing based on transversal filter [118–120]. It works by generating weighted and delayed copies of the real-time RF signal in the optical domain and then combing them during photo-detection. Transversal filters can achieve arbitrary RF transfer functions by changing the weight of taps. The taps usually are offered by multi-wavelength source such as discrete laser arrays [121–123], Bragg grating arrays [124, 125], electro-optical generated combs [16], or mode-locked fiber lasers [126]. However, taps offered by these approaches are limited in quantity and the complexity. Compared to the methods mentioned above, integrated OFC based on micro-ring resonators have distinct advantages in offering multi-wavelength sources with smaller in size, more conducive to integration and achieve larger free spectrum range (FSR) [127–132].

In 2015, Thach for the first time combined an OFC based on micro-ring with a RF photonic transversal filter to realize a 20-tap Hilbert transformer, and we called this system "the real-time signal processing system", leading the research upsurge of RF photonic fil-ter based on OFC [23]. The real-time signal processing system proposed in this paper consists of three main sections: OFC generation, comb shaping, and RF photonic filtering. In the generation of OFC, pump wave generated by a continuous-wave tunable laser is amplified by a high power Erbium-doped fiber amplifier (EDFA1) and then the pump wave enters into the micro-ring resonator and generate OFC. In the second section of the RF photonic signal processing system, a reconfigurable filter (WaveShaper) shapes the comb lines of OFC according to the required tap coefficients. Afterwards, the comb lines of OFC as a multiple-wavelength source enter into the last section of this system with going through a 2x2 balanced Mach-Zehnder modulator for modulating comb lines to realize negative and positive tap coefficients, 2.122 km single-mode fiber (SMF) for delaying the different filter taps, and a second fiber amplifier (EDFA2) for compensating for loss and separating the comb. The optical signals are finally detected by photodiodes to output RF signals. The real-time signal processing system's frequency response is measured with an RF vector network analyzer (VNA). Afterwards, this real-time signal processing system also realized a reconfigurable RF photonic intensity differentiator [49] including the first, second and third order intensity differentiator and a reconfigurable true RF time delay capable of yielding a phased array antenna [49]. The first multi-channel RF tunable microwave true time delay lines for phased array antenna based on this real-time signal processing system has higher performance and lower size. By shaping the comb lines of OFC, this system realized the antenna which achieved high angular resolution, wide range of beam steering angles, and small variation in beam steering angle. In [106], the photodetector of the system was replaced by balanced photodetector which consisted of two mutually matched and consistent photodiodes. Two photodiodes could differentially subtract the current signal and the signal was amplified and output. Compared with previous photodetector, balanced photodetector could effectively counteract the influence of both light source and environmental noise, thereby enhancing the signal-to-noise ratio. In this work, Sim not only realized a significantly finer 50 GHz FSR offering up to 80 comb lines based on a Kerr micro-comb but also realized a fractional Hilbert transformer with arbitrary fractional orders. In [49], a broadband Local Oscillator (LO)-free photonic RF mixer was also proposed based on this real-time signal processing system which could convert RF frequency to the U-band. The comb lines offered by OFC served as a RF local oscillator for RF signal conversion. Finally, this work realized a low ratio of -6.8 dB between output RF power and IF power and the spurious suppression of signal was up to 43.5 dB. Following this mixer, Xingyuan Xu el at. further proposed a novel approach to Optical artificial networks (ONN) based on this real-time signal processing system [107]. ONN is a novel approach to neural network computation that utilizes principles and technologies of photonics [133, 134]. In traditional electronic neural networks, information transmission occurs through electronic signals processed and transmitted on silicon chips [135–137]. In contrast, optical neural networks leverage the properties of light for information transmission and processing, potentially offering advantages in speed and energy efficiency [138–143].

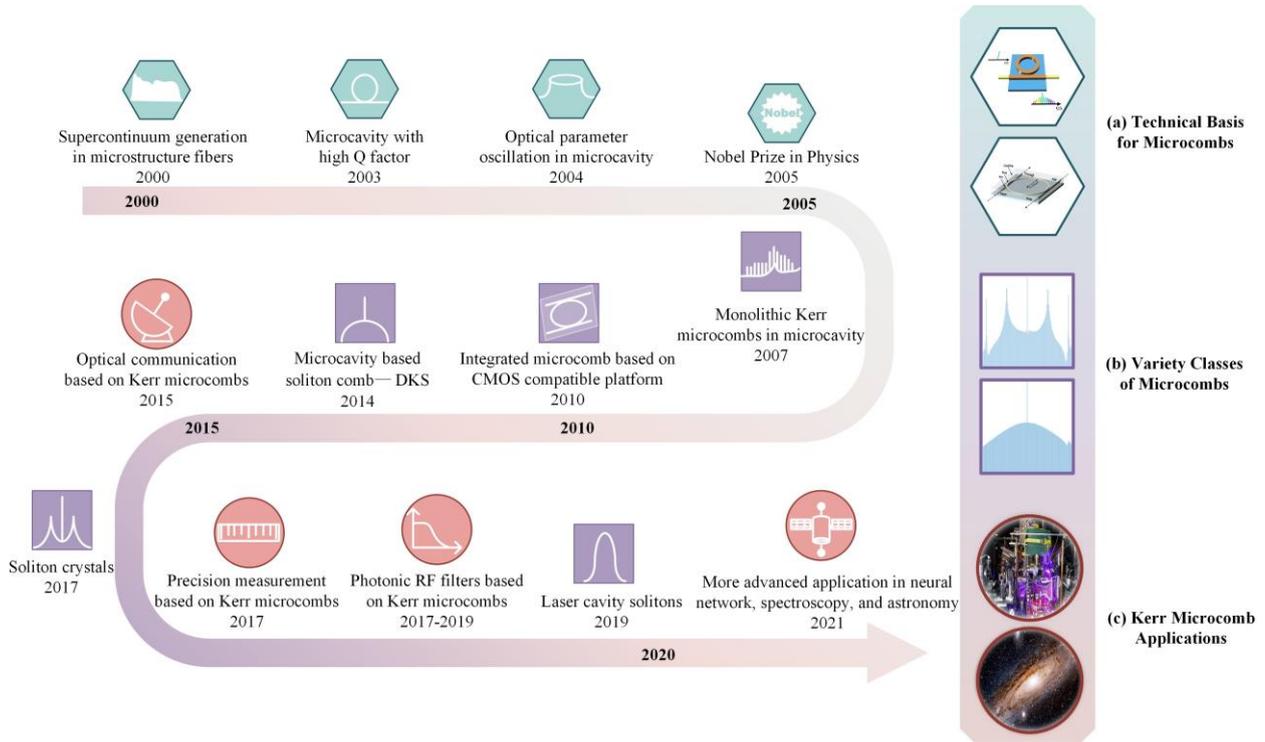

**Fig. 4:** Timeline for the development of Kerr microcombs. (a) Technical basis for microcombs [76–80]. (b) Variety classes of microcombs generated in the development process [20, 22, 81–84]. (c) Applications of the Kerr microcombs [73, 85–91].

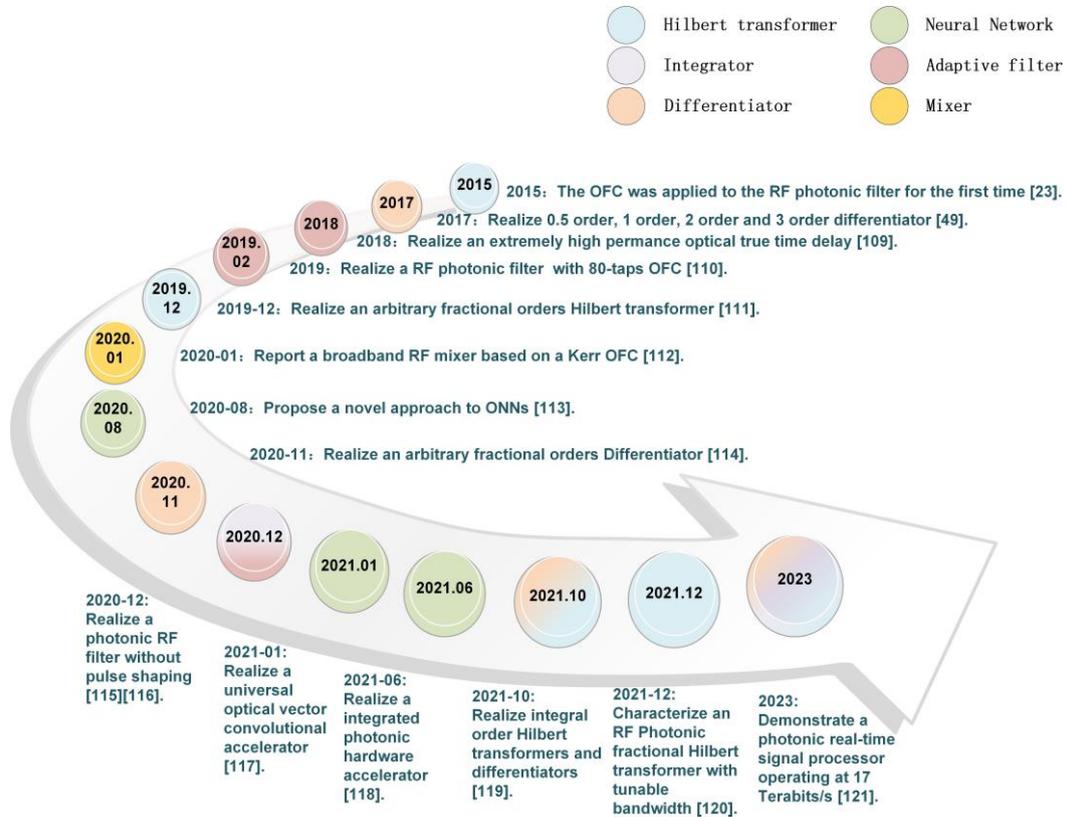

**Fig. 5:** Timeline for the development of real-time signal processing based on OFC [23, 49, 104–114].

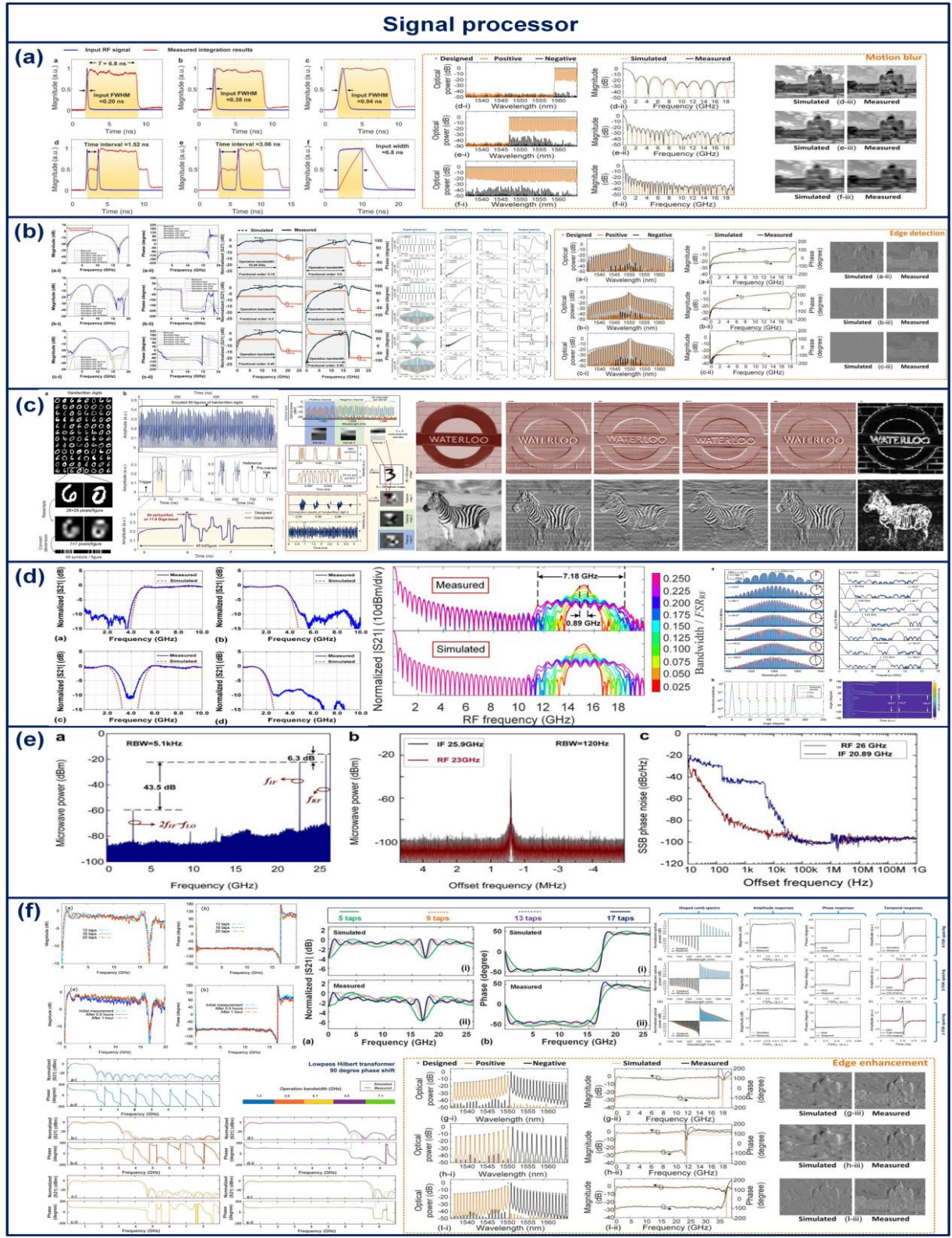

**Fig. 6:** The results of real-time signal processing system for different applications based on OFC. (a) Integrator [109, 114]. (b) Differentiator [49, 108, 112, 114]. (c) Neural network [104, 107, 111]. (d) Adaptive filter [49, 105] . (e) Mixer [49]. (f) Hilbert transformer [23, 106, 112–114].

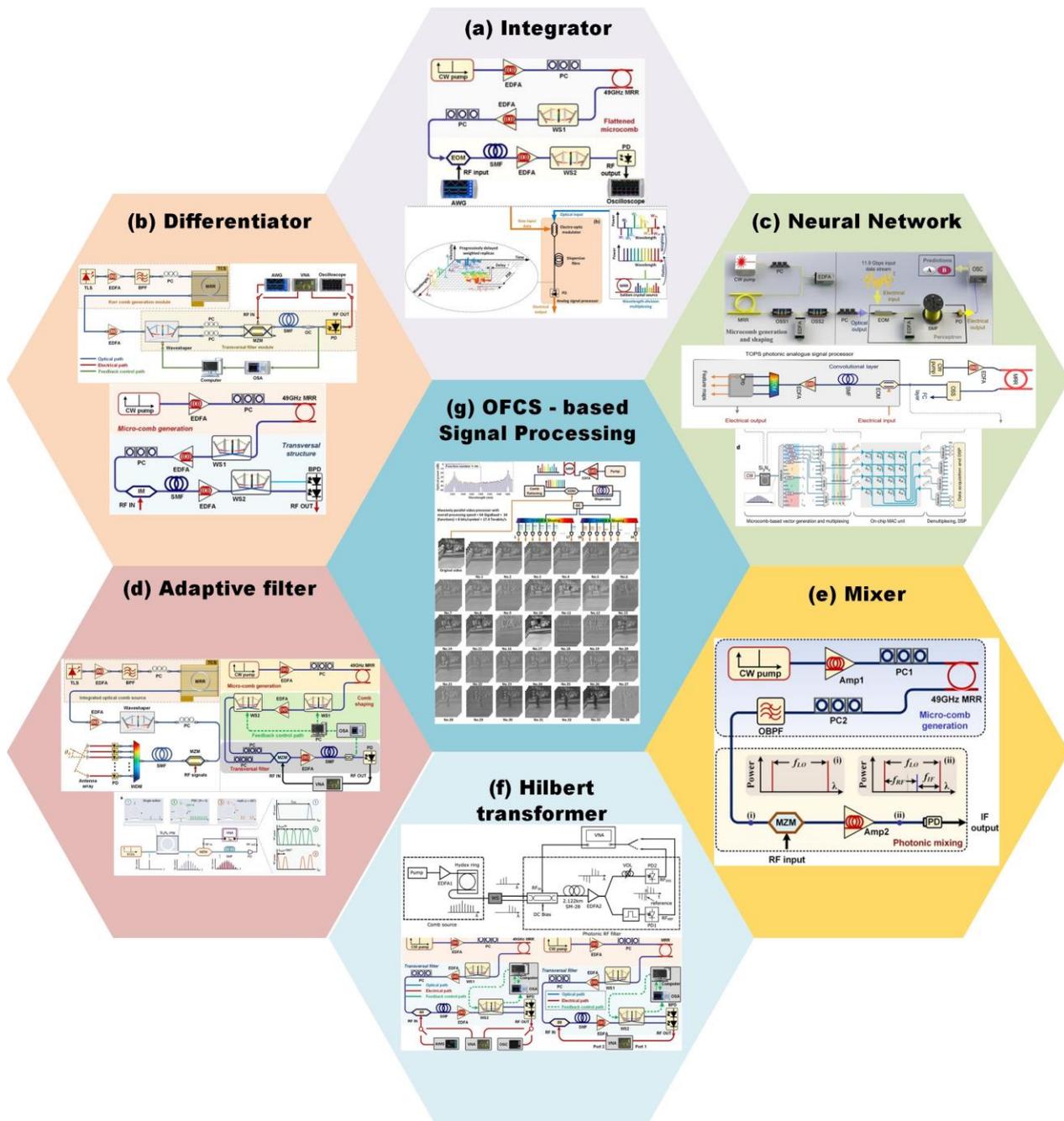

**Fig. 7:** different structure of real-time signal processing system for different applications based on OFC. (a) Integrator [109, 114]. (b) Differentiator [49, 108, 112, 114]. (c) Neural network [104, 107, 111]. (d) Adaptive filter [49, 105] . (e) Mixer [49]. (f) Hilbert transformer [23, 106, 112–114].

The basic idea behind ONN is to use light's properties for information transmission and processing between neurons. Typically composed of components such as light sources, modulators, amplifiers, and detectors, optical neural networks simulate the functionality of neurons and process information through optical signals.

In this paper, a new approach based on the real-time signal processing system including OFC utilized wavelength, time, spatial, multiplexing to compute vector dot products, which could use straightforward flatting to convert dot products into vectors for performing matrix operation. The novel ONN finally achieved a record throughput of 95.2 Gbps per unit and then it was applied to the standard benchmark tests including handwritten digits and predicting benign/maglignant cancer classes, capable of achieving >93% accuracy and >85% accuracy, respectively. In 2020, a novel differentiator was demonstrated based on this real-time signal processing system [108]. In contrast to previous differentiator [49], this differentiator not only operated directly on the real-time RF signal rather than the optical domain but also was based on a novel the form of OFC named "soliton crystals", thus effectively realized reconfigurable arbitrary fractional orders ranging from 0.15 to 0.9. At the same time, Kippenberg's team also designed an RF photonic filter, but the signal processing system they constructed was slightly different from this real-time signal processing system in this paper [110]. The signal processing system They designed didn't have a WaveShaper and adjusted the OFC spacing by triggering a perfect soliton crystals.

A novel photonic integrator based on the real-time signal processing system with transversal structures was reported [109], which offered multiple wavelength channels because of OFC so that each path could be controlled independently and thus improving high reconfigurability and accuracy. What's more, in contrast to previous real-time signal processing system, the comb shaping approach of this system was optimized from optical power shaping to impulse response shaping in order to reduce errors. This photonic integrator was capable of offering a large integration time window of 6.8 ns and a time resolution of 84 ps.

After optimizing this system, Xu proposed a universal optical vector convolutional accelerator using the same hardware of real-time signal processing system, which could operate at more than 10 TOPS and generate a 250,000 pixels images convolutions [104]. At the same year, Kippenberg et al. also proposed a novel integrated photonic accelerator involved in ONN based on soliton microcombs and was capable of operating multiply-accumulate at speed of 1012 MAC operations per second [111].

In 2021, after demonstrating fractional order differentiator, Sim Tan proposed a integral order differentiator including 1st, 2nd and 3rd order three kinds based on this real-time signal processing system [112]. Soon after, Sim Tan demonstrated an configurable RF photonic Hilbert transformer with tunable bandwidths as well as centre frequency on the basis of the same system [113]. This Hilbert transformer could achieve a tunable bandwidth ranging from 1.2 GHz to 15.3 GHz as well as a tunable centre frequency from baseband to 9.5 GHz by adjusting the weight of taps and programming. Following this, Sim Tan further first proposed a video images processor based on this real-time signal processing system which was highly reconfigurable by programmable control [114]. This video images processor could perform not only different functions such as integral and fractional order differentiation, fractional order Hilbert transforms, and integration but also images processing approaches including edge detection, motion blur and edge enhancement without changing the physical hardware as show in Fig. 7 (g). This work demonstrated that this video images processor could simultaneously process over 399,061 real-time signals with an ultrahigh bandwidths of 17 Tbs/s. In addition, This result confirmed that the theory was in good agreement with the measurement, and laid a foundation for the integration of ultrahigh-bandwidth real-time signal processing system.

## 5 Discussion

We review the development of real-time signal processing system based on Kerr OFC in chronological order in Fig. 5. The first part reviews the development process of RF photonic signal processing methods. The second part introduces the classification and development trend of OFC in detail. In the final part, we further review the development of real-time signal processing system based on OFC and the different parallel functions implemented in chronological order. Thereby, we review the timeline of this real-time signal processing system and we operated a series of optimization for this system.

The real-time signal processing system realized many parallel functions such as integrator, differentiator, neural network, adaptive filter, mixer and Hilbert transformer showcased in Fig. 6 (a)-(f) and Fig. 7 (a)-(g). For Hilbert transformer and differentiator, the number of wavelength determines the number of comb lines while the number of taps in turn directly determines the performance of this system. Arbitrary fractional and integral orders differentiators and tunable-bandwidth Hilbert transformers were proposed based on this real-time signal processing system with excellent performance.

For the system mentioned in this article, the performance of the optical frequency comb has reached a high state, and many record-setting functions have been completed including that the optical vector convolutional accelerator, and the new world record (up to 17 Terabits/s) photonic real-time signal processor. Nonetheless, there are still many applications as well as more parallel functions that ultimately could be implemented based on this system.

In terms of the comb lines shaping, tap errors during comb shaping can lead to deviations between experimental results and theory, as well as non-ideal impulse responses in the system. These errors stem from various sources such as optical micro-comb instability, waveshaper inaccuracies, wavelength-dependent gain variation in optical amplifiers, chirp induced by optical modulators, and third-order dispersion in dispersive fibers. To mitigate these issues, real-time feedback control loops can be employed. Initially, the power of comb lines is detected by an optical spectrum analyzer (OSA) and compared with ideal tap weights to generate an error signal. This signal is then fed back into the waveshaper for system calibration, enhancing comb shaping accuracy. Alternatively, to further refine tap errors and improve accuracy, the feedback loop's error signal can be derived directly from the measured impulse response instead of raw optical power. This involves measuring RF Gaussian pulse replicas across all wavelengths to obtain the system's impulse response, from which peak intensities are extracted to determine RF-to-RF wavelength channel weights accurately. Subsequently, the extracted channel weights are compared with desired weights to derive an error signal for programming the loss of waveshaper. Through multiple iterations of this comb shaping loop, an accurate impulse response is achieved, compensating for system non-idealities and enhancing the accuracy of RF photonic transversal filter-based signal processors. Importantly, this calibration process is conducted only once.

As for the integration of real-time signal processing system, the potential of this system is significant due to its integration capabilities with current nanofabrication techniques [144]. The microcomb source is already integrated and fabricated using CMOS-compatible processes [145, 146]. The several key components of this system have been successfully integrated using cutting-edge nanofabrication techniques [147–149]. These include the optical pump source [145, 146], optical spectral shapers [147], $LiNO_3$ modulators [149], dispersion media [148], and photodetectors. Furthermore, advanced integrated microcombs have demonstrated reliable generation of soliton crystals in operational settings [146]. Monolithic integration of the entire real-time signal processing system promises enhanced performance, compactness, and energy efficiency. Despite lacking complete integration, employing discrete integrated OFC instead of laser arrays already provides significant advantages for RF systems in terms of performance, size, cost, and complexity.

Recent advancements have shown that soliton crystals can achieve various RF functions without requiring spectral shaping, solely through different pumping conditions that produce varied spectra [110]. This expands the capabilities of OFC beyond what is achievable with DKS states. Such developments are particularly advantageous as they eliminate the need for spectral shapers, components in RF systems that ultimately require integration efforts.

## Acknowledgments

This work was supported by the Australian Research Council (ARC) Centre of Excellence in Optical Microcombs for Breakthrough Science "COMBS" (No. CE230100006), and by the National Natural Science Foundation of China (No.62293495 and No.62201021) and the Young Elite Scientists Sponsor- ship Program (No.2023QNRC001).